**"You should see my partners' fingers": A Qualitative Study of Construction Artisans' Perspective on Technical Innovation**

Mariavittoria Masotina[1], Laura Brogioni[2], Roberto Paladini[3], Patrik Pluchino[1,2], Anna Spagnolli,[1,2] Luciano Gamberini[1,2]

[1]*Human Inspired Technologies Research Centre, University of Padova, via Venezia 8, 35131, Padova, Italy*

[2]*Dept. of General Psychology, University of Padova, via Venezia 8, 35131, Padova, Italy*

[3]*IUAV University of Venice, Santa Croce, 191, 30135 Venice, Italy*

**Corresponding author**

Anna Spagnolli, Via Venezia 8, 35131 Padova, Italy, anna.spagnolli@unipd.it

# ABSTRACT

Construction industry scholars have advocated increasing digitalization as a harbinger of manifold improvements, from safe training to efficient waste management. Small construction enterprises, which often face greater difficulties in embracing such a paradigm, are frequently overlooked in investigations of stakeholders' views on technical innovation. This study aims to start filling this gap by investigating the views of small construction enterprises on technical innovation.

We report the themes that emerged from the qualitative analysis of interviews with construction artisans in North-East Italy in 2018-2019, including installers, restorers, carpenters, painters, and upholsterers (N=25). We asked what makes new technical devices acceptable to them and conducted inductive and deductive thematic analyses to identify recurrent arguments supporting their positions. The analysis identified fifteen premises underlying the interviewees' positions on technical innovation, grouped around four issues: Is technological innovation part of my job? Is the financial cost of innovating worthwhile? Will the new technology be practical? What part of my business is served by communication devices?

The study sheds more light on an underrepresented sector and the articulated view of technological innovation in small enterprises, where there is no distance between investors and users. It also adopts a qualitative methodology that can be exported to investigate the acceptability requirements of other samples and specific devices.

**Keywords:** Technical innovation; Construction sector; Small enterprises; Thematic Analysis; Artisans; Reasons.

## INTRODUCTION

According to the latest and most extensive public consultation ever conducted by the European Union, finding digital solutions for the construction sector is one of the priorities and societal challenges for the next ten years (Directorate-General for Research and Innovation, 2023). Digitalization could help reduce some structural problems in the construction industry: the coordination of professional and non-professional stakeholders and expertise (architects, contractors, subcontractors, customers, suppliers) (Merschbrock and Munkvold, 2015); the search for a talented workforce (Maskuriy *et al.*, 2019); the safety training and monitoring via virtual reality and wearable technologies, e.g.,(Guo, Li and Li, 2013); and the production of waste via strategic project management, e.g., (Tang, Cass and Mukherjee, 2013). Digitalization could reduce the costs of labor-intensive processes and materials by using embedded sensors automatically tracking the equipment (Sardroud, 2012) and by delivering on-time, on-budget construction projects (Lawson, Ogden and Goodier, 2014), (Maskuriy *et al.*, 2019). Data analysis tools could support the managers' decisions (Bilal *et al.*, 2016). Simulation technologies could reduce errors in the early stage of construction, e.g., Building Information Modeling (BIM)(Lee, Park and Won, 2012), and show customers detailed, vivid renditions of a planned building, e.g., Augmented/Virtual Reality) (Hussien *et al.*, 2019).

However, the construction sector seems to steer clear of this innovation paradigm (Muñoz-La Rivera *et al.*, 2021). Innovation sets challenges that are hard to meet, such as motivating employees to learn new processes (Craveiroa *et al.*, 2019), changing their internal standards (Sony and Naik, 2019), and needing maintenance and training (Xu and Duan, 2019). Tight schedules encourage the use of the same technologies across projects and limit the possibility of experimenting with new ones (Turk, 2021). Small and medium-sized enterprises are more likely to suffer from these difficulties than larger companies (Nowotarski and Paslawski, 2017). A study on the agriculture sector in Germany, for example, emphasized how small and medium farmers are afraid that digitalization would bring hidden costs (i.e., education and office work), replace their trust net via digitally-mediated trust tools (e.g., blockchain) and allow big suppliers and buyers adjust prices by accessing their business data (Linsner *et al.*, 2019) (Linsner *et al.*, 2021).

Taking into account companies' difficulties and reservations is core to a 'sustainable, human-centric and resilient' approach to industrial innovation (Directorate-General for Research and Innovation, 2022). While the previous innovation paradigm, called Industry 4.0, focused on digitization and automation to improve the industry's efficiency, timeliness, and flexibility (Lu, 2017) (Müller, Kiel and Voigt, 2018), a human-centric approach to industrial innovation prioritizes a rewarding and motivating working environment that genuinely responds to the workers' needs (European Commission, 2021). This approach resonates with the view of technology as symbolic artifacts, defined by meanings and values as much as by technical functionalities (Button, 1993); (Gagliardi, 1990); (Joyce *et al.*, 2023); (Kling, 1980); (Zucchermaglio, Bagnara and Stucky, 1995).

Our goal in the study reported here is to focus on micro and artisanal enterprises in the construction sector and understand their view on adopting technical innovations in their work. We study a sample of artisans in the Venetian area, where artisanship is intensely represented, and adopt a qualitative approach, shifting the focus from technology *acceptance* to *acceptability*.

## TECHNOLOGY ADOPTION IN THE CONSTRUCTION SECTOR

A few studies investigated the factors affecting technology adoption in the construction sector. Chung and colleagues (Chung *et al.*, 2008) surveyed 281 users of Enterprise Resource Planning (ERP) software. They identified several factors that significantly affect the intention to use ERP systems. In particular, perceived usefulness proved to be a highly significant predictor, influenced by subjective norms, perceived ease of use, the quality of system output, alignment with the company's essential business functions, and the tangibility of results. Accordingly, they recommended designing the ERP system to cover the company's essential business functions, be easy to use, encourage all enterprise members to use it, and describe clear outputs and positive results.

Davies and Harty (Davies and Harty, 2013) analyzed the opinions of 762 construction employees on Building Information Modeling systems (BIMs). They concluded that BIM's perceived usefulness depended on respondents' perceptions of the organization's support for using the system and on the system's compatibility with existing working procedures. Consequently, the authors recommended 'balancing messages about the transformative nature of BIM with reassurance about compatibility, relevance, and continuity for individual-level acceptance.'

Lee, Yu, and Jeong (Lee and Yu, 2016) proposed a BIM Acceptance Model (BAM) and ran a comparative study between Korean (N = 114) and American (N = 50) experienced BIM users. In the American sub-sample, they obtained similar results to Davis and Harty (Davies and Harty, 2013): BIM's perceived usefulness depended on the belief of being in an organization open to technological innovations that will support them in using the system; it also depended on BIM's compatibility with the potential adopters' previous experience, work practice, system use, and needs. In addition, they found an effect of 'personal competency' (i.e., personal innovativeness and willingness to try out any new information technology) on perceived usefulness and the social and market environment within and outside the organization.

Sepasgozaar, Shirowzhan, and Wang (Sepasgozaar, Shirowzhan and Wang, 2017) studied the intention to use a scanner for the 3D acquisition of the buildings' models. Based on five case studies, the authors suggested that the technology's perceived usefulness (i.e., performance expectancy) and ease of use (i.e., effort expectancy) can influence users' intentions, sense of efficacy, and perceived support during the implementation phase.

Finally, an investigation of the reasons for adopting technology was carried out in Northern Italy by Bettiol et al. (Bettiol, Maria and Capestro, 2018). They surveyed 668 manufacturing enterprises, including small ones. They also relied on pre-formulated answer options, which prevented unexpected arguments from emerging. They proposed a pre-defined list of technological systems, some of which might not have been known to the participants (i.e., laser cutting, robotics, big data/cloud computing, additive manufacturing, Internet of Things, 3D scanners, and augmented reality).

This body of literature identifies some key factors facilitating the adoption of new technologies in the construction sector, formulated in the very abstract terms typical of TAM models (Venkatesh *et al.*, 2003): perceived ease of use, usefulness, subjective norms, and demonstrability of the technology results; in turn, these factors are affected by contextual variables such as compatibility with existing procedures, market environment and availability of organization support, and personality variables such as personal innovativeness and sense of efficacy.

## STUDY APPROACH: FROM ACCEPTANCE TO ACCEPTABILITY

The factors highlighted by the studies described in the previous section are theoretical, high-level categories investigated with questionnaires whose formulation remains unaltered across devices and studies to allow standardization. They follow the approach of the Technology Acceptance Model (TAM) (Davis, 1985), (Davis, Bagozzi and Warshaw, 1989), which can be traced back to social psychology and, more specifically, to some classic models of attitudes elaborated by Ajzen and Fishbein (Theory of Planned Behavior, TPB, Ajzen, 1991). TAM has been extended to incorporate other factors to become the Unified Theory of Acceptance and Use of Technology (UTAUT, (Venkatesh *et al.*, 2003) and has been applied to several contexts, including the construction sector of interest here.

The problem with this model is that its findings are too abstract and generic to be actionable. For instance, knowing that a given technology is perceived as not easy to use does not end the investigation with implementable advice. Further investigation would be needed to identify which specific features are responsible for the low usability. Also, the way the issue of technology adoption is framed seems unbalanced in favor of the manufacturer, making the final user appear capricious and unpredictable, ultimately forced to accept the technology. A more human-centered approach to technology adoption is achieved by shifting the focus from changing users' intentions to improving the technology's alignment with users' priorities. The output is not a single number expressing users' level of technology acceptance, but a list of explicit, specific reasons that can make the technological innovation acceptable or not. We use the term "acceptability" rather than "acceptance" to mark this different approach, although the two terms are sometimes treated as synonyms in the literature (Alexandre *et al.*, 2018).

To identify this list of reasons, we followed an argumentative approach. According to psychologists Mercier and Sperber (Mercier and Sperber, 2011), human reasoning is an exquisitely argumentative process that does not pursue the truth but rather conclusions supported by convincing arguments. These arguments are partly based on evidence and partly on culturally shared and accepted commonplaces (Billig, 1996) and are formed when the person is trying to articulate a position. Hence, we framed the acceptability of technology as controversial, i.e., as an issue that can inspire opposing positions (van Eemeren and Grootendorst, 2004), and identified the arguments participants set forth to articulate such positions.

## METHOD

### *4.1 Procedure and data*

The study took place in Italy in 2018-2019, where many construction firms are small artisan firms, and their owner is simultaneously the entrepreneur, the investor, and one of the workers[1]. They provide the specific resources needed by a larger company, with which they join in temporary consortia. The larger company acts as a general contractor (Confartigianato Imprese, 2015). We focused on Venice and Padova, lively areas for small artisanal enterprises.

[1] The average size of enterprises in Italy is about 60% of that in other EU countries, while micro-enterprises with fewer than 10 employees, accounting for 95.4% (or 3.8 million units) of the total, represent 47% of employment

The participants were contacted through two artisans' associations. The interview occurred at the artisans' businesses or, if this was not possible for organizational reasons, at the union's office. All participants were asked to read and sign an informed consent form before starting data collection, and could request any clarification they needed. Then they completed a questionnaire and were interviewed. All interviews were audio-recorded and later transcribed.

The types of data collected in the study are synthesized in Table 1. Basic socio-demographic information was collected with an *ad hoc* questionnaire. Participants' positions on adopting technical innovations in their jobs were collected during semi-structured interviews, which allowed the interviewees to articulate their positions in their own words (Mann, 2016). The interviews started with two warm-up questions and continued with questions about the advantages, disadvantages, and facilitatory conditions of adopting technical innovations in their job. The answers could be followed by requests for clarification, such as 'Why?' or 'What do you mean?,' consistent with the semi-structured interview format.

Table 1. Data collection protocol

| | |
|---|---|
| **Questionnaire** | • Socio-demographic information (Participant's age, gender, and nationality)<br>• Work sector and size (in number of employees) of the participant's business<br>• Usage of digital technologies for specific activities (e-commerce, e-governance, e-banking, electronic incoming, marketing online, certified e-mail) and processes (i.e., planning, production, stock management, and administration; purchase, sales, marketing, and customer services).<br>• Technologies to communicate with customers and suppliers (i.e., phone calls, e-mails, and dedicated online services). |
| **Interview warm-up** | • What is your business?<br>• Which computerized technology do you use in your work? |
| **Interview questions** | • In your opinion, which factors would facilitate the introduction of new technologies in your business? |

compared to 21% (or 2.1 million units) in Germany, 22% in France and 27% in the United Kingdom. Italian SMEs tend to be more focused on manufacturing and construction than in the rest of Europe: SMEs operating in this industry contribute 31% to the total value added produced (EU-27 average: 21%) and 25% of jobs in SMEs (EU-27 average: 20%, Source Econ, 2019).

| | |
|---|---|
| | • What could be the possible advantages of introducing new technologies in your sector?<br>• What could be the possible disadvantages of introducing new technologies in your sector?<br>• Are there other activities we have not mentioned yet that can benefit from introducing new technologies? |

### *4.2 Analysis*

All interviews were transcribed and then analyzed with a reflexive thematic approach. A theme is a repeated pattern of meaning found in the data. In our case, the themes were the premises underlying the interviewees' position on adopting or not adopting technologies at work. Thematic analysis is a qualitative procedure for identifying, analyzing, and reporting themes across a dataset (Braun and Clarke, 2006). A reflexive thematic analysis is one in which themes are generated during the analysis (Braun, Clarke and Gray, 2017).

The analytic procedure included an inductive and a deductive phase and involved three different researchers. The inductive phase consists of detailed readings of the transcripts to derive themes that summarize the data and are linked to the research questions (Thomas, 2006). Two of the authors worked together to identify the passages in which the interviewees expressed a position in favor of or against adopting technical innovations in their job; a passage was, for instance, 'Let us say that new technologies can be very attractive, a nice website, a nice proposal, for customers who may not know us.' These passages are called argumentation tokens. Then, they identified the token's premises; for instance, the token mentioned above rested on the premise that 'technology is useful for attracting new customers.' This step was interpretative because the premise was often implicit. Eventually, 35 premises were identified across the whole dataset, accounting for all argumentative tokens in the interviews and concerning the reasons for accepting or rejecting technology. The premises were finally grouped by similarity to reduce overlaps and redundancy. Fifteen premises resulted from this process.

The deductive phase of the analysis began with these 15 premises and used them as categories to code all tokens; this analysis tested whether the categories saturated the data (Braun and Clarke, 2021). Two coders were involved; one participated in the first inductive phase, and the other was a newly recruited coder who could bring a fresh pair of eyes to the process. They worked independently to code all tokens, then compared the results and found that 60% of the tokens were coded identically. Finally, they cooperated to (a) reduce the overlap in the premises' definition, (b) make the premises' description more straightforward, and (c) achieve a shared decision about divergent coding. This agreement-seeking process is a standard procedure in a reflexive thematic analysis and contributes to making the interpretation intersubjective (Bae, 2017). At the end of this process, we obtained 15 finalized premises.

### *4.3 Ethics*

The study did not include deceitful, risky procedures or collect sensitive data. The study followed the international standard code of conduct for research with humans (American Psychological Association, 2017) and the General Data Protection Regulation (European Union, 2016), as the data collection occurred within the European Union. Every participant signed an informed consent explaining the study's aim, methods, and the participants' rights. They had the right to withdraw from the study at any time without providing any justification. The participation was voluntary and did not involve financial compensation. The participants were granted anonymity by assigning each one a sequential ID. All information provided during the interviews that could suggest the identity of participants or their collaborators (e.g., suppliers' names) was anonymized during transcription.

### *4.4 Participants*

The participants were 25 artisans (23 males and two females). According to Guest, Bunce, and Johnson (Guest, Bunce and Johnson, 2006), a sample size like ours suffices to achieve saturation in the case of research on a clear domain with which all participants are familiar, with narrow objectives, and when participants are interviewed independently from one another and following the same protocol. The participants' ages ranged from 29 to 79 years (M = 49.68, SD = 12.25), and they were owners of their businesses (except for two participants, who were employees); they were based in the province of Venice (11 participants) and the nearby province of Padova (14 participants). All of them worked in the construction sector, with different roles: electrical systems (N = 5), plumbing systems (N = 4), restoration (N = 5), joinery (N = 4), painting (N = 4), tiling (N = 2), and upholstery (N = 1). They worked in micro-businesses composed solely of themselves (N = 6) or with 2 to 9 employees (Md = 3). Almost all participants (N = 21) dealt with private customers, while a minority (N = 4) also served the civil and industrial sectors.

In terms of levels of digitization, the computerized services our participants reported using were banking, certified e-mail, invoicing, and marketing (Figure 1). The processes supported by computerized systems are production, stock management, and administration (N = 15); planning (N = 14); and purchase, sales, marketing, and customer services (N = 15). Communication with customers and suppliers was mostly via phone (with customers: N = 25; with suppliers: N = 25) and e-mail (with customers: N = 25; with suppliers: N = 20).

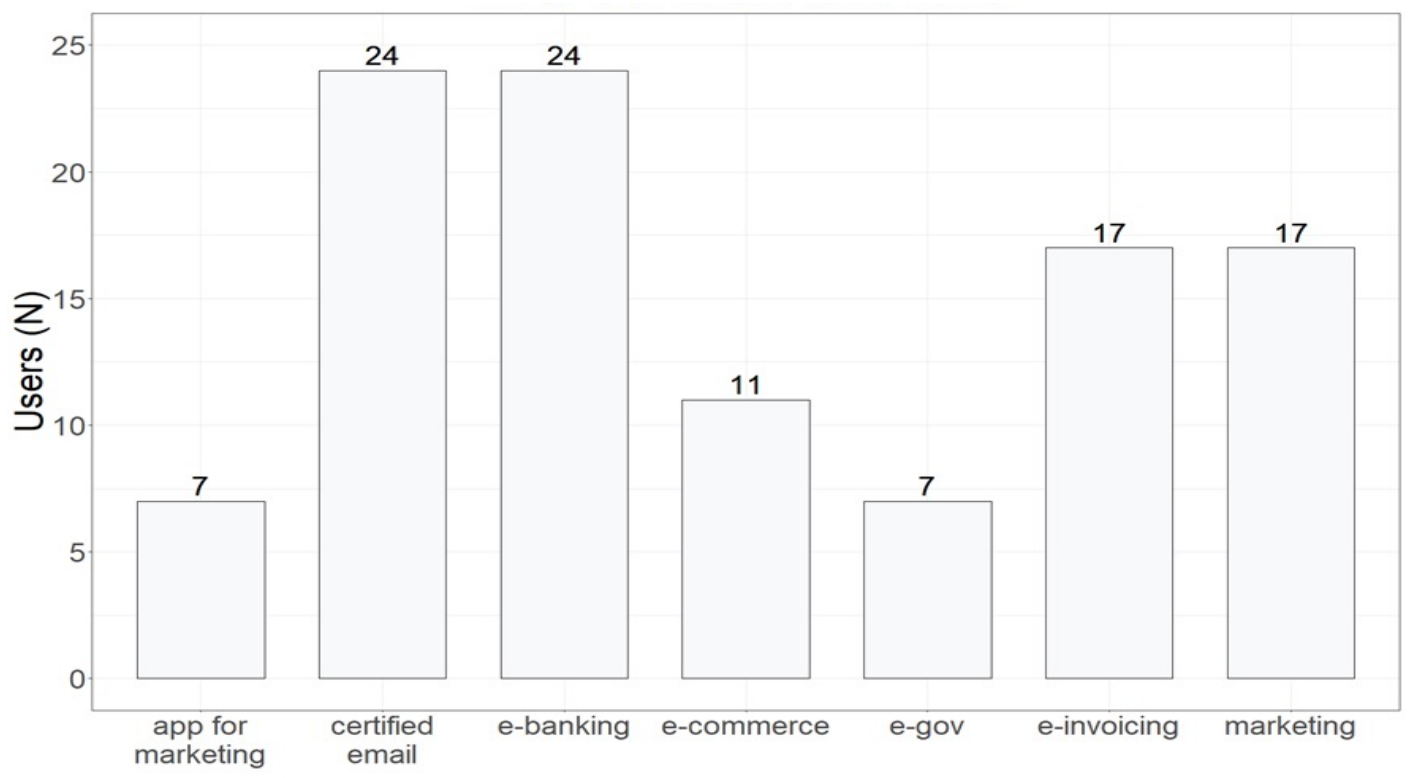

**Fig. 1** Participants' use of different categories of digitalized services.

## RESULTS

After analyzing the interviews, we found fifteen premises underlying the interviewees' positions on adopting technical innovations in their jobs. We grouped them based on four controversies: Is technological innovation part of my job? Is the financial cost of innovating worthwhile? Will the new technology be practical? What part of my business is served by communication devices? Within each controversy, the premises are listed alphabetically, accompanied by the frequency in the sample (i.e., the number of participants whose answers were associated with a given premise), and exemplified by a few excerpts. The frequency is not a claim about the relative importance of each premise but an index of its recurrence, and it is reported for completeness.

### *5.1 Is technological innovation part of my job?*

*Irreplaceability (15 participants):* Unlike other sectors of the economy, the artisans we interviewed did not feel threatened by technological innovation, as they seemed to believe that no technology would ever replace them.

> A little bit of technology could only relieve fatigue, but it could not cancel human labor. I don't see competition from technology at this moment - P03.

> Our work is manual; artisanal work is physical. About measuring, I see how they [technologies] can help. But no technology can help me make less effort, finish sooner, or work better - P22.
> We work in an artisanal way, where the variants are always very complicated, and there are not many innovations for putting a stucco - P15.

To them, the role of technology is to assist human artisans who are responsible for the relevant part of the work.

> Developing a telescopic meter to upload the numbers directly to the tablet would be very nice because I wouldn't have the reading error. I mean, you may read 523 and write 532. It's stupid, but it's 9 mm of difference. So, that would be very interesting because it reduces errors to a minimum - P06.

*Consistency (19 participants):* Our interviewees were favorable about adopting new technical solutions and devices if they served tasks pertinent to their work and clients.

> We chose not to adopt rendering because, in our opinion, this kind of transaction [with customers] pertains to the architect. We are the makers - P01.
> I have loyal customers; my client roster is small. I have no reason to enlarge it - P13.
> You can send the documentation by e-mail to 50% of customers; the other 50% are elderly and won't use it or know how to do it - P25.

This view extended to claiming the incompatibility between automation and their work processes:

> They [the artisan's products] are like clothes for different people; clothes cut for one person would hardly fit another. I mean, it's a unique product... it is not replicated - P13.

*Natural updating (23 participants):* A large part of the sample stated that technological innovation was necessary for the survival of artisans' work. Hence, resisting innovation per se would be nothing more than laziness or bad practice.

> It is not that we are willing; it is that either you change or... bye-bye - P08.
> Almost every [tool] has an app that connects it to the mobile phone. If we used to do it by hand, now there is an app on the mobile phone - P11.
> One uses the tools one knows, and out of laziness, I would say, one wouldn't venture into other things - P19.

### *5.2 Is the financial cost of innovating worthwhile?*

*Reducing personnel cost (15 participants):* The interviewees seemed favorable to technologies that reduced the number of people needed to do a job.

> I'd have to either work or take care of it [new management software]. I can't do both. So, I need to work manually to earn my daily bread, and I have no one to do this job for me. And, in any case, there are no financial possibilities to pay someone to do it for me - P17.
> We are developing a tablet system where the [technicians] will fill in all the needed paperwork. Currently, this work is done in the office, [in the future] they will do it directly. […] If we have to be three people to do a job, we will have no margin. But, if only one person is needed, this changes everything; we will become even more competitive in the market - P04.

*Short-term return (13 participants):* Technical innovation was appreciated if associated with immediate financial benefits and affordability.

> We have been considering this laser machine for two years, but we just got back on our feet after building the warehouse - P14.
> I mean, at the end of the month, there should be a financial return associated with the initial investment - P13.
> It would be really convenient if we could reduce those profit losses caused by any wrong count of hours - P08.

The pursuit of profit also needs to realistically consider the customers' financial means to avoid scenarios like the following:

> Good morning, Mrs. Maria. Just because I am here with all this stuff [fancy machinery] attached, you must give me 500 euros, okay? Then, for the repair itself, it's 3 euros - P13.

*Time efficiency (21 participants):* Technologies were appreciated if they reduced the time necessary to complete a specific task.

> [We need] increasingly advanced equipment to improve service and reduce times... because [...] you have to increase productivity; otherwise, you are out. So, if ten years ago a technician did the job in one day, now he does it in half a day – P04.
> Now we were considering a single system, […] to eliminate as many steps as possible. Because let's say we are generally in the pipeline: you make your report,

> then you bring it here, you take it back, and then you still have to compare it with what you foresaw in the estimate. Therefore, there are a lot of steps taking time, and we want to eliminate all, well, a good part of these steps – P08.

### *5.3 Would the new technology suit the circumstances of my job?*

*Ease of use (17 participants):* Our interviewees viewed new technologies positively when they simplified the work process and were easy to use.

> Measuring the height of a house with a laser becomes ... easier than ... with a telescopic meter - P03.
> Technology is the computer I use to get my quotations, to check a website of interest, and WhatsApp to send messages. This is what I know. I know how they work because I use them, and maybe I don't know a whole world of other technologies, but I don't even have much time to get to know them - P07.

*Flexibility (14 participants):* The interviewees appreciated that the same technology could be used across different situations, tasks, and stakeholders.

> [In our software], you manage deadlines, presentations, and everything. It's all wrapped up in the same program - P14.
> The technicians use the same management software we use in the office, but in the technical version, so they will only see the information they are interested in and their appointments - P04.
> The day before yesterday, I sent an AutoCAD 3D file to make the legs of a metal table. [My collaborator] replied: 'Please send it to me in pdf because I can't read it - P01.

*Physical conditions (6 participants):* Technology was appreciated if it suited the physical conditions of the workplace

> No, the computer stays at home. Otherwise, it ends up badly. (...) Well, when you go around in our work, you put it in a construction site, dust and all. You don't want that - P20.
> You should see my partners' fingers… this is the problem! - P16.

*Reliability (19 participants):* A precondition to accepting a technological solution in the workplace was its reliability.

> When I have to ask for an estimate, I always use e-mail. Because, in the end… it's more tangible [while] by phone calls… I call you, and you write a note that gets lost. By e-mail, it's all written - P01.
> Even if it's all online, we print it anyway. […] A couple of months ago, there was a blackout. My colleague and I twiddled our thumbs for two hours - P02.
> We use the laser meter a lot, especially for the measurements of the mosquito nets that must be precise. Because you can't be wrong; if you're wrong, you throw it away - P21.

### *5.4 Would any part of my business use innovation?*

*Collaboration (21 participants):* An appreciated characteristic of a technical solution was its integration with the technologies adopted by collaborators, suppliers, and customers. This led to a preference for elementary collaboration tools.

> WhatsApp helped us a lot in managing the issues… because a picture is worth a thousand words - P06.
> Everyone had their mail, their calendar station, and their contacts. We decided to delete this thing and turn it into [name] so that we had shared calendars and contacts. This change simplified things, let's say, because [in the past] if we had to do a meeting, we had to say: 'you are there, you are not there .'Now, we look, and we know - P01.

*Handling vast amounts of data (12 participants):* Innovative technology is useful when it simplifies the management and analysis of relatively large amounts of data compared with traditional, paper-based management tools.

> I assume that everyone has [management software], especially installers. Maybe not all of them, but the slightly bigger installers use management software, at least for invoicing, because otherwise, it becomes [impossible]. Also, there are so many variations of [product] code that it has become impossible to manage them all - P10.
> My storage management is my brain because I have just a few things - P09.
> If I have a project or something that consists of x pages, I would need 4 or 5 monitors, so… - P13.

*Human connection (12 participants):* Artisans valued human contact with collaborators, customers, and suppliers; face-to-face interaction was often preferred.

> We work in people's homes, so we need to have a certain relationship with them. [We need] friendship, almost an informal relationship with our customers - P25.
> He's the typical customer who needs some presence, some contact; he is not the customer you attract through an advertisement. […] In my opinion, this is the most loyal type of customer because they look for a more human relationship, not just for the price difference - P07.
> In the evening, the boys [the employees] all come back here [at the headquarters] because they have to bring the vehicles back. They stop here, and we talk about the work issues. [This is a crucial moment] because you can understand how the work went; if one is more nervous because maybe he had some difficulties, then you try to give him a hand – P02.
> I go there because I often need to go there and talk directly with [my suppliers] because they can give you helpful advice. [..] I saw that this kind of relationship - the human relationship - provides you with more information - P16.

*Access to information (18 participants):* The interviewees seemed to value technology that removed the obstacles to accessing work-related information.

> [After having] acquired data on the construction site, the reports can be done directly from a tablet or smartphone to simplify it [the process]. It can also be done during the journey from the construction site to the home - P10.
> [Using my management software], when something happens, they [my employees] already have the product, they know how much it costs. Therefore, they can give an estimate to the customer and tell him [sic] 'Look, to repair your machine, you will need x and x, equal money x' - P04.
> Those with more technological power can answer better; they can solve the problem better- P01.

*Promotion (21 participants):* New technologies were appreciated by artisans interested in customers who cannot be reached by word of mouth.

> We have seen that a satisfied customer sends us at least five more customers. Word of mouth is very important. In recent years a website has also been developed, so by seeing the site, maybe someone… - P14.
> It's also a matter of image. Because if somebody goes to work with a mallet and a chisel rather than a computer, one has a slightly different image - P04.

> Well, if you find a skeptic [customer], you can upload the file with the design of the arched window he has chosen, and you can show him on the computer how it opens. He changes his face...- P06.
> But the color must be quite truthful. Because: what happens? It happens that when you show [that paint], and the customer falls in love, and then when it's spread, it is a whole other thing... This is a serious problem - P09.

## DISCUSSION

The analysis of the interviews suggests that the construction artisans participating in our study do not object to adopting technical innovations in their jobs per se. Although the construction industry's culture might seem resistant to change and innovation (Oesterreich and Teuteberg, 2016), our results remind us that technology is an integral part of construction workers' everyday activities, and that innovation that facilitates their work is welcome. Indeed, a large share of the construction industry's investment is in machinery and equipment (54,28%) (Directorate-General for Research and Innovation, 2022). The initial questionnaire shows some computerized technical innovations that artisans have already integrated into their work routines. Resistance to change seems, then, like an oversimplification, preventing a more articulated view of the reasons for making technical innovation acceptable. Moreover, the interviewees do not seem to perceive that their job is threatened by technical innovation like in other sectors in the manufacturing or creative industry, where the worker's concerns about being replaced by machines have been reported (e.g., (Jacob *et al.*, 2023); (Spagnolli *et al.*, 2020). Perhaps the projects of small construction companies, 'time-limited, site-based unique (Oesterreich and Teuteberg, 2016), p.123)' is a protection factor against this risk: the kind of tailored craftsmanship provided by construction artisans is not easily automated; the clients' buildings (e.g., renovations) vary greatly in age, conditions, and size, as do the circumstances of their work.

The literature on technology acceptance suggests that perceived usefulness and ease of use are pivotal factors in forming an intention to use new technologies (Venkatesh *et al.*, 2003). Our results corroborate this finding but also identify additional requirements, such as flexibility, reliability, and robustness. The interviewees emphasized the need for a device that is easy to use, flexible across sites, reliable, and robust enough to withstand the physical conditions of the workplace, including dust and harsh temperatures. It must be noted that in microenterprises, where the owner is also one of the workers, acceptability criteria connected to user satisfaction with the product are more influential than in larger companies, where purchase decisions are made by managers who will never use the equipment first-hand. The small size of these companies also explains another acceptability factor: the innovation must provide quick financial returns; small companies would not have the financial backing to afford unprofitable purchases. Finally, a technical innovation must be straightforwardly relevant to their business: for example, CAD software is perceived as a tool for architects, not construction workers.

Technology's compatibility with the users' work is pinpointed as an essential condition for its acceptance in the literature on construction workers (Davies and Harty, 2013); (Lee and Yu, 2016); to this requirement, our interviews add that any new device must be compatible with the (simple) communication tools already used by suppliers and customers. Unlike a previous study by Bettiol (Bettiol, Maria and Capestro, 2018) addressing companies of various sizes, the small companies we interviewed did not mention restoring and maintaining international competitiveness and made-in-Italy production; perhaps the reason is that our interviewees operate with customers located nearby and rely on larger contractors to find distant customers.

On a more conceptual note, some of the premises we found point to the relevance of work and of a positive identity (e.g., (Dutton, Roberts and Bednar, 2010); (Walsh and Gordon, 2008), which connect the beliefs of the individual worker with their work environment, its values, and moral standards. In TAM models, instead, the social context is an external factor influencing the individual's choice (Davis, 1985); (Davis, Bagozzi and Warshaw, 1989); (Venkatesh *et al.*, 2003), and a specific psychological construct is missing that connects the individual with their work context.

The study described in this paper presents some limitations. First, our participants worked in the same geographical area (Northeast of Italy). Although the study was conducted in the Veneto region of Northern Italy, construction enterprises are not specific to that area, so the results can be compared with those of construction artisans in other regions.

Second, male workers represented almost all our interviewees. Although this reflects the gender gap in the construction industry, this limitation results in an underrepresentation of female workers' perspectives.

## CONCLUSION

This research aimed to identify the peculiar characteristics of artisanal construction micro-enterprises. We aimed to illustrate how their views articulate, rather than reduce them to either pro- or anti-innovation. We identified 15 premises around which their views are built, partly consonant with other reports and partly new or more specific, as prior studies mostly adopted a quantitative approach, measuring very abstract constructs. We hope that this study sheds light on a sector that is seldom investigated with respect to its view on technical innovation and provides a viable methodological approach to other studies with similar goals.

**FUNDING**

This work was partially supported by a research agreement of the HIT Research centre at the University of Padova with small artisan enterprises in the Venetian Area within the project "Audit tecnologico e orientamento sull'uso delle tecnologie per l'ottimizzazione del lavoro all'interno dei cantieri edili." The funding source had no role in the study design, data collection, analysis and interpretation, report writing, or the decision to submit the article for publication.